\documentclass[nofootinbib,prd]{revtex4}
\pdfoutput=1

\usepackage{times,epsfig}
\usepackage{graphics,graphicx}
\usepackage{amsmath}
\usepackage{amsfonts}
\usepackage{epsfig}
\usepackage{color}

\usepackage{enumerate}

\usepackage{mathrsfs}
\usepackage{amssymb}
\usepackage{bm}

\newcommand{\aj}{AJ}

\def\lsim{\,\lower2truept\hbox{${<\atop\hbox{\raise4truept\hbox{$\sim$}}}$}\,}
\def\gsim{\,\lower2truept\hbox{${> \atop\hbox{\raise4truept\hbox{$\sim$}}}$}\,}
\def\simlt{\mathrel{\rlap{\lower 3pt\hbox{$\sim$}}
        \raise 2.0pt\hbox{$<$}}}
\def\simgt{\mathrel{\rlap{\lower 3pt\hbox{$\sim$}}
        \raise 2.0pt\hbox{$>$}}}

\newcommand{\fnlt}{$f_{\rm NL}$}

\def\be{\begin{equation}}
\def\ee{\end{equation}}
\def\ba{\begin{eqnarray}}
\def\ea{\end{eqnarray}}

\def\hksqrt{\mathpalette\DHLhksqrt}
\def\DHLhksqrt#1#2{\setbox0=\hbox{$#1\sqrt{#2\,}$}\dimen0=\ht0
\advance\dimen0-0.2\ht0
\setbox2=\hbox{\vrule height\ht0 depth -\dimen0}%
{\box0\lower0.4pt\box2}}

\begin{document}

\title{Probing primordial non-Gaussianity via iSW measurements with SKA continuum surveys}

\author{Alvise Raccanelli$^{1,2}$, Olivier Dor\'{e}$^{1,2}$, David J. Bacon$^{3}$, Roy Maartens$^{3,4}$, Mario G. Santos$^{4,5}$, Stefano Camera$^{6}$, Tamara M. Davis$^{7,8}$, Michael J. Drinkwater$^{7}$, Matt Jarvis$^{4,9}$, Ray Norris$^{10}$, David Parkinson$^{7}$ \\~}

\affiliation{
$^{1}$ Jet Propulsion Laboratory, California Institute of Technology, Pasadena, California 91109, USA \\
$^{2}$ California Institute of Technology, Pasadena, California 91125, USA \\
$^{3}$ Institute of Cosmology \& Gravitation, University of Portsmouth, Portsmouth P01 3FX, UK \\
$^{4}$ Physics Department, University of the Western Cape, Cape Town 7535, South Africa \\
$^{5}$ SKA SA, 3rd Floor, The Park, Park Road, Pinelands, 7405, South Africa\\
$^{6}$ CENTRA, Instituto Superior T\'{e}cnico, Universidade de Lisboa, Portugal\\
$^{7}$ School of Mathematics \& Physics, University of Queensland, Brisbane QLD 4072, Australia \\
$^{8}$ ARC Centre for All-sky Astrophysics (CAASTRO), Australia \\
$^{9}$ Astrophysics, Department of Physics, University of Oxford, Oxford OX1 3RH, UK \\
$^{10}$ CSIRO Astronomy \& Space Science, Epping NSW 1710, Australia
}

\date{}

\begin{abstract}
The Planck CMB experiment has delivered the best constraints so far on primordial non-Gaussianity, ruling out early-Universe models of inflation that generate large non-Gaussianity. Although small improvements in the CMB constraints are expected, the next frontier of precision will come from future large-scale surveys of the galaxy distribution. The advantage of such surveys is that they can measure many more modes than the CMB -- in particular, forthcoming radio surveys with the Square Kilometre Array will cover huge volumes.
Radio continuum surveys deliver the largest volumes, but with the disadvantage of no redshift information. In order to mitigate this, we use two additional observables. First, the integrated Sachs-Wolfe effect -- the cross-correlation of the radio number counts with the CMB temperature anisotropies -- helps to reduce systematics on the large scales that are sensitive to non-Gaussianity. Second, optical data allows for cross-identification in order to gain some redshift information. We show that, while the single redshift bin case can provide a $\sigma(f_{\rm NL}) \sim 20$, and is therefore not competitive with current and future constraints on non-Gaussianity, a tomographic analysis could improve the constraints by an order of magnitude, even with only two redshift bins.
A huge improvement is provided by the addition of high-redshift sources, so having cross-ID for high-z galaxies and an even higher-z radio tail is key to enabling very precise measurements of $f_{\rm NL}$.
We use Fisher matrix forecasts to predict the constraining power in the case of no redshift information and the case where cross-ID allows a tomographic analysis, and we show that the constraints do not improve much with 3 or more bins.
Our results show that SKA continuum surveys could provide constraints competitive with CMB and forthcoming optical surveys, potentially allowing a measurement of  $\sigma(f_{\rm NL}) \sim 1$ to be made. Moreover, these measurements would act as a useful check of results obtained with other probes at other redshift ranges with other methods.
\end{abstract}


\date{\today}

\maketitle


\section{Introduction}
\label{sec:intro}
Cosmology with radio surveys is attracting growing attention (see e.g.~\cite{masui10,raccanelliradio,ansari11,pritchard11, camera12,mao12,maartens12,masui12,battye12,hall13,rubart13,patel13,camera13,ferramacho14,metcalf14,bharadwaj11,bharadwaj13,wolz}), partly driven by the prospect of the Square Kilometre Array (SKA)\footnote{www.skatelescope.org}, which promises to deliver huge-volume surveys of the galaxy distribution, using the H{\sc i}-21cm emission or the radio continuum emission of galaxies.
Among the biggest challenges in cosmology is to constrain observables that might distinguish between different models of inflation, in particular via non-Gaussianity in the primordial perturbations.
Measuring primordial non-Gaussianity (PNG) is key to understanding the physics of the early universe. The current highest precision measurement is from the Planck CMB experiment~\cite{planckfnl}: 
\be\label{pfnl}
f_{\rm NL}=3.5 \pm 7.5~(1\sigma),
\ee
where $f_{\rm NL}$ is the local PNG parameter (corresponding to squeezed states in the bispectrum). This result has effectively ruled out models of inflation that generate a large local PNG. Note that we use the large-scale structure convention: $f_{\rm NL}^{\rm LSS} \approx 1.3 f_{\rm NL}^{\rm CMB}$~\cite{xia10ng}.

It is important to have independent measurements of $f_{\rm NL}$ from galaxy surveys, as the improvements in precision from future Planck data and other CMB experiments are not expected to significantly improve on the current $\sigma_{f_{\rm NL}}\sim 7$. Since 3-dimensional galaxy surveys can measure many more modes than the 2-dimensional CMB, these surveys will deliver the next level of precision on \fnlt. Currently the best constraints from (optical) galaxy surveys~\cite{ross13,giannantonio13,Ho13} are still well behind Planck, with $\sigma_{f_{\rm NL}}> 20$ and dominated by systematic errors. The next-generation {\em Euclid} galaxy survey\footnote{www.euclid-ec.org} in the optical/infrared is predicted~\cite{euclid} to match the constraining power of the CMB experiments, with $\sigma_{f_{\rm NL}}\sim 4$. An SKA HI survey in intensity mapping out to redshift $z\sim 3.5$, could achieve~\cite{camera13} $\sigma_{f_{\rm NL}}\sim 1$, considering statistical errors only.

Radio continuum surveys can cover 30,000 deg$^2$ out to high redshift, giving the largest-volume surveys of the galaxy distribution. However their constraining power is weakened by the lack of redshift information. In~\cite{ferramacho14}, this weakness was counter-balanced by separating out different types of radio galaxies, allowing the use of the multi-tracer method~\cite{seljak08, mcdonald} to beat down cosmic variance, and resulting in much improved constraints on \fnlt. Here we adopt a different approach, using a combination of two methods to mitigate the problem of redshifts:
\begin{itemize}
\item 
Cross-correlation of the radio counts with CMB temperature anisotropies, i.e. the integrated Sachs-Wolfe (iSW) effect, allows us to reduce systematics on large scales, where there is maximum sensitivity to PNG.
\item
Cross-identification of radio sources with optical data in order to split the radio sources into redshift bins. 
\end{itemize}
We show that these methods allow an SKA continuum survey to match or improve the level of constraining power predicted for future optical surveys such as Euclid.

The paper is organized as follows.
In Section~\ref{sec:isw} we discuss the integrated Sachs-Wolfe effect, and in Section~\ref{sec:ng} we briefly review  non-Gaussianity.
Section~\ref{sec:surveys} presents the properties of the SKA and pathfinder surveys and the data they should collect.
Section~\ref{sec:fisher} describes the methodology used for obtaining the results, which are shown in Section~\ref{sec:results} and finally discussed in Section~\ref{sec:conclusions}.


\section{Integrated Sachs-Wolfe effect}
\label{sec:isw}

The iSW effect~\cite{sachs67} is a gravitational shift due to the evolution of the gravitational potential as photons pass through matter under- or over-densities in their path from the last scattering surface to us. In an Einstein-de Sitter universe, the blueshift of a photon falling into a well is cancelled by the redshift as it climbs out, giving a zero effect.
In the presence of a time variation of the local gravitational potential, due to a dark energy component or a deviation from Einstein's theory of gravity, however, potential wells are modified and photons will experience a blue- or red-shift, leading to a net change in photon temperature which accumulates along the photon path. This translates into CMB temperature anisotropies proportional to the variation of the gravitational potential.

The iSW effect can contribute significantly to the CMB temperature fluctuations on large angular scales, but it enhances only the low $\ell$ multipoles, and is smaller than other CMB anisotropies. Typically the iSW signal is weak and hard to interpret due to large cosmic variance, and cross-correlation with tracers of the density field is necessary to bring it to measurable levels. Here we are interested in using the iSW to improve the signal of PNG in the galaxy counts. The CMB temperature anisotropies are not sensitive to \fnlt\, at first order. Rather, we use the iSW to reduce systematics on the very large scales that are most sensitive to PNG. 

We can write the cross-correlation power spectrum between the surface density fluctuations of galaxies and CMB temperature fluctuations as (for more details, see e.g.~\cite{crittenden96, nolta04, cabre07, raccanelli08}):
\begin{equation}\label{eq:ClgT}
C_{\ell}^{gT} = \langle a_{{\ell}m}^g a_{{\ell}m}^{T*} \rangle = 4 \pi
\int \frac{dk}{k} \Delta^2(k) W_{\ell}^g(k)
W_{\ell}^T(k),
\end{equation}
where $W_{\ell}^g$ and $W_{\ell}^T$ are the galaxy and CMB window functions, respectively, 
and  $\Delta^2(k)$ is the logarithmic matter power spectrum today.

The galaxy window function can be written as:
\begin{equation}\label{eq:flg}
W_{\ell}^g(k) = \int dz\, \frac{dN}{dz} b(z) D(z,k) j_{\ell}[k\chi(z)] ,
\end{equation}
where  $(dN/dz)dz$ is the mean number of sources per steradian with redshift $z$ within $dz$, brighter than the flux limit; $b(z)$ is the bias factor relating the source to the mass overdensity, $D(z,k)$ is the linear growth factor of mass fluctuations; $j_{\ell}$ is the spherical Bessel function of order $\ell$, and $\chi(z)$ is the comoving distance. 

The window function for the iSW effect is:
\begin{equation}\label{eq:flT}
W_{\ell}^T(k) = \frac{3}{2} \Omega_{m} \left(\frac{H_0}{ck}\right)^2 \int dz \left[\dot{\Phi}(z,k) +\dot{\Psi}(z,k) \right] j_{\ell}[k\chi(z)],
\end{equation}
where $ \Phi,\Psi$ are the gravitational potentials. For the concordance model and more generally for standard dark energy models in general relativity, we assume a scale-independent growth factor, $D=D(z)$, and equality of the potentials, $\Psi=\Phi$.

Equations \eqref{eq:ClgT}--\eqref{eq:flT} show that the cross-correlation between the CMB and the galaxies depends on the clustering and bias of tracers of the large-scale structure, and on the evolution of the gravitational potentials. 
For this reason it has been used to test and constrain cosmological models such as the evolution and clustering of structures ~\cite{raccanelli08, massardi10}, models of dark energy~\cite{xia09} and alternative cosmological models~\cite{giannantonio08dgp, bertacca11}.

Cosmic magnification~\cite{loverde07}, redshift-space distortions~\cite{rassat09} and general relativistic effects on very large scales~\cite{maartens12}, will affect measurements of the cross-correlations, and so in principle we should include those corrections into our modeling. However, while necessary when doing a proper comparison with data and important for measuring other cosmological parameters, they are small effects that will not affect the conclusions of this work, so we will not include them in this paper.

Additionally, as shown in~\cite{liu11}, including polarization information to the CMB and radio continuum cross-correlation studies reduces the uncertainties by 5 to 12\%. We do not include this information, but it is worth keeping in mind that with this addition, which will also be available from the surveys that we  consider, the errors on the parameters may reduce further.


\section{Primordial non-Gaussianity}
\label{sec:ng}
The amplitude and shape of clustering on large scales, described by the angular power spectrum, can provide important cosmological information.
Constraining PNG offers a powerful test of the generation mechanism for cosmological perturbations in the early Universe. While standard single-field models of slow-roll inflation lead to very small departures from Gaussianity, non-standard scenarios (such as e.g. multi-field inflation) allow for a larger level of non-Gaussianity (see e.g.~\cite{bartolo04, komatsu10, wands10}).
Probing PNG can be done using the CMB (see~\cite{planckfnl} and references therein) or 
the large-scale structure of the Universe (see e.g.~\cite{matarrese00, dalal08, slosar08, desjacques10, xia10ng, RaccanellifNL, Camera14}).

For the local or squeezed limit of the bispectrum, the gauge-invariant Bardeen potential is given as a perturbative correction to a Gaussian random field $\phi$:
\begin{equation}
\label{eq:fnl}
\Phi=\phi+f_{\rm NL}\left(\phi^2-\langle\phi^2\rangle\right) .
\end{equation}
To constrain PNG from large-scale structure surveys, we exploit the fact that a non-zero $f_{\rm NL}$ in Equation~\eqref{eq:fnl} introduces a scale-dependent modification of the large-scale halo bias (see e.g.~\cite{dalal08}): 
\begin{equation}
\label{eq:ng-bias}
b(z,k)=\bar{b}(z)+\Delta b(z, k) =\bar{b}(z)+ [\bar{b}(z)-1] f_{\rm NL}\delta_{\rm ec} \frac{3 \Omega_{m}H_0^2}{c^2k^2T(k)D(z,k)}, 
\end{equation}
where $\bar{b}(z)$ is the usual bias calculated assuming Gaussian initial conditions (which is assumed to be scale-independent, as we look only at large scales), $T(k)$ is the transfer function ($\to 1$ on large scales) and $\delta_{\rm ec}$ ($\approx 1.45$) is the critical value of the matter overdensity for ellipsoidal collapse.

\begin{figure*}[htb!]
\includegraphics[width=0.6\linewidth]{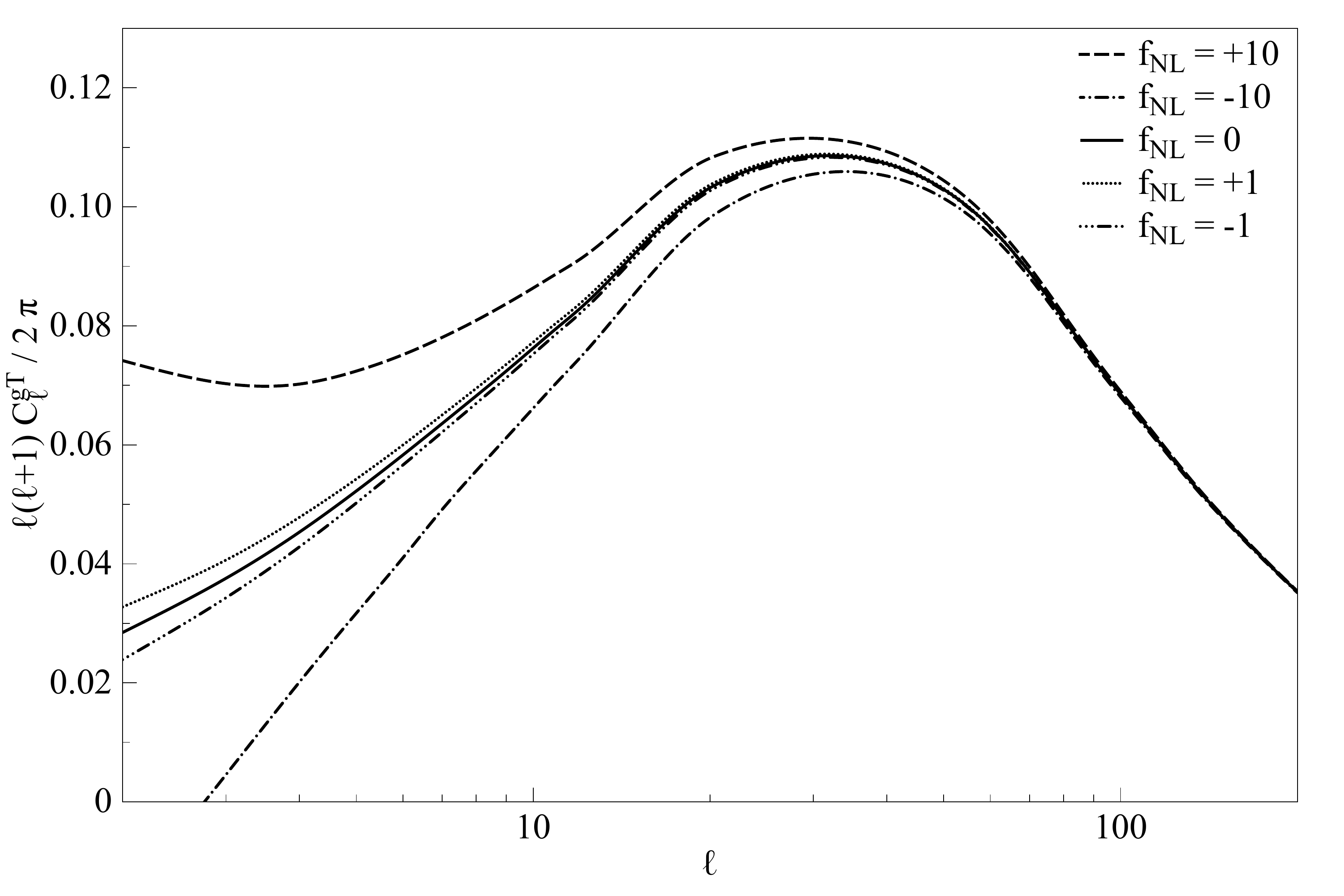}
\caption{Modifications to the cross-correlation of CMB with radio sources, for different values of \fnlt.
}
\label{fig:nGiSW}
\end{figure*}

The Planck best-fit value for \fnlt\, in \eqref{pfnl} already takes into account large-scale second-order effects near recombination that can bias the measurement of the primordial signal. In addition, various projection effects from looking down the past lightcone can also bias the large-scale PNG signal and these are also incorporated in the result \eqref{pfnl}.  

In the case of large-scale structure observations, there are analogous effects that can bias a measurement of the primordial signal. A second-order relativistic correction to the Poisson equation introduces an effective $f_{\rm NL}^{\rm eff} \approx-2.2$~\cite{Bartolo:2005xa, Verde:2009hy, Camera14}, which needs to be added to the observed value.

This affects the best-fit value but should not affect the errors that we calculate.
There are further relativistic projection effects at first order, following from the fact that we observe galaxies on the past lightcone and not on a constant-time surface~\cite{Yoo:2010ni,Bonvin:2011bg,Challinor:2011bk, Bertacca:2012tp}.
These effects make a growing contribution as $k\to 0$, which can contaminate the signal of local PNG, as shown by~\cite{Bruni:2011ta} (see also \cite{Jeong:2011as,Yoo:2011zc,raccanelli13}). They do not have the same scale and redshift dependence as local non-Gaussianity, and so there is no simple description in terms of an effective parameter $f_{\rm NL}^{\rm eff}$. However the amplitude of the effects is typically comparable to $|f_{\rm NL}| =O(1)$. 
An accurate measurement of PNG will need to remove contamination due to these effects, but a careful treatment of this is beyond the scope of the present work.


\section{SKA and its Pathfinders}
\label{sec:surveys}

In order to forecast the iSW constraints we will consider several large sky continuum surveys that will be provided by the SKA telescope and its pathfinders.
\begin{itemize}
\item
{\bf ASKAP}~\cite{johnston08}: The final ASKAP (Australian SKA pathfinder) configuration will consist of 36 12m dishes spread over a region of 6 km in diameter; it will have a $30 \, \rm deg^2$ instantaneous field of view, which will provide a large survey speed. EMU (Evolutionary Map of the Universe~\cite{emu}) is the 3$\pi$~sr continuum survey planned for ASKAP. It will make a deep, radio continuum survey of the entire Southern Sky, extending as far North as $+30 \deg$. EMU should be 45 times more sensitive, and with an angular resolution 5 times better than NVSS~\cite{condon98}.

\item
{\bf MeerKAT}~\cite{jonas09}: The South African SKA pathfinder will consist of 64 dishes of 13.5m each spread over a region 8~km in diameter. Although no large sky survey is planned at the moment, for the sake of completeness we consider here the possibility of a 30,000 $\rm deg^2$ survey done in 10,000 hours (same as EMU). With an instantaneous field of view of $\sim 1\rm deg^2$, a bandwidth from 900MHz to 1670MHz and a sensitivity, A/T, of 300 m$^2$/K, this will translate into a rms of $\sim 10\mu$Jy per pointing. We will therefore speak about the MeerKAT survey or EMU interchangeably.

As representative of the precursor surveys we therefore consider a $3\pi$ sr survey of the sky ($\sim 30,000 \rm deg^2$) to a rms of $\sim 10\mu \rm Jy$.

\item
{\bf SKA}: The SKA will be developed in two stages. The first stage currently encompasses two mid-frequency facilities (~GHz) operating within Australia and South Africa. Given that they have similar continuum survey speeds at around 1GHz we consider them interchangeably, although we note that the phased array feed technology on the Australian antennae may provide a higher survey speed at $\nu > 1$~GHz, whereas below this frequency the single-pixel feeds on the South African site may survey the sky more rapidly. Sensitivities for SKA1 for a survey over 30,000 deg$^2$ with 10,000 hours are expected to be around 3 $\mu \rm Jy$ rms depending on the specific requirements and final setup. In the second stage of the SKA, the plan is to extend the array by a factor of 10, thus significantly increasing the survey power of the facility. Final sensitivities for SKA2 with the same survey type could range between 300 nJy to 50 nJy rms depending on the telescope field of view.
Since no definitive survey plan for the SKA in either phase has been made, we consider a range of surveys that seem feasible as we move into phase 1 and beyond this into phase 2.
We consider three surveys with rms noise of $1, 2.5, 5 \mu \rm \, Jy$ compared to the value of $10\mu \rm \, Jy$ assumed for the pathfinders case. Although a more futuristic SKA2 will in principle go below the 1 $\mu \rm \, Jy$ case, we took the more conservative approach of limiting the analysis to the cases above.
\end{itemize}

In all cases we assume a $3\pi$ sr survey of the sky and a detection threshold of 5$\sigma$.


\subsection{Galaxy Distribution and bias}
\label{sec:galdist}
\begin{figure*}[htb!]
\includegraphics[width=0.7\linewidth]{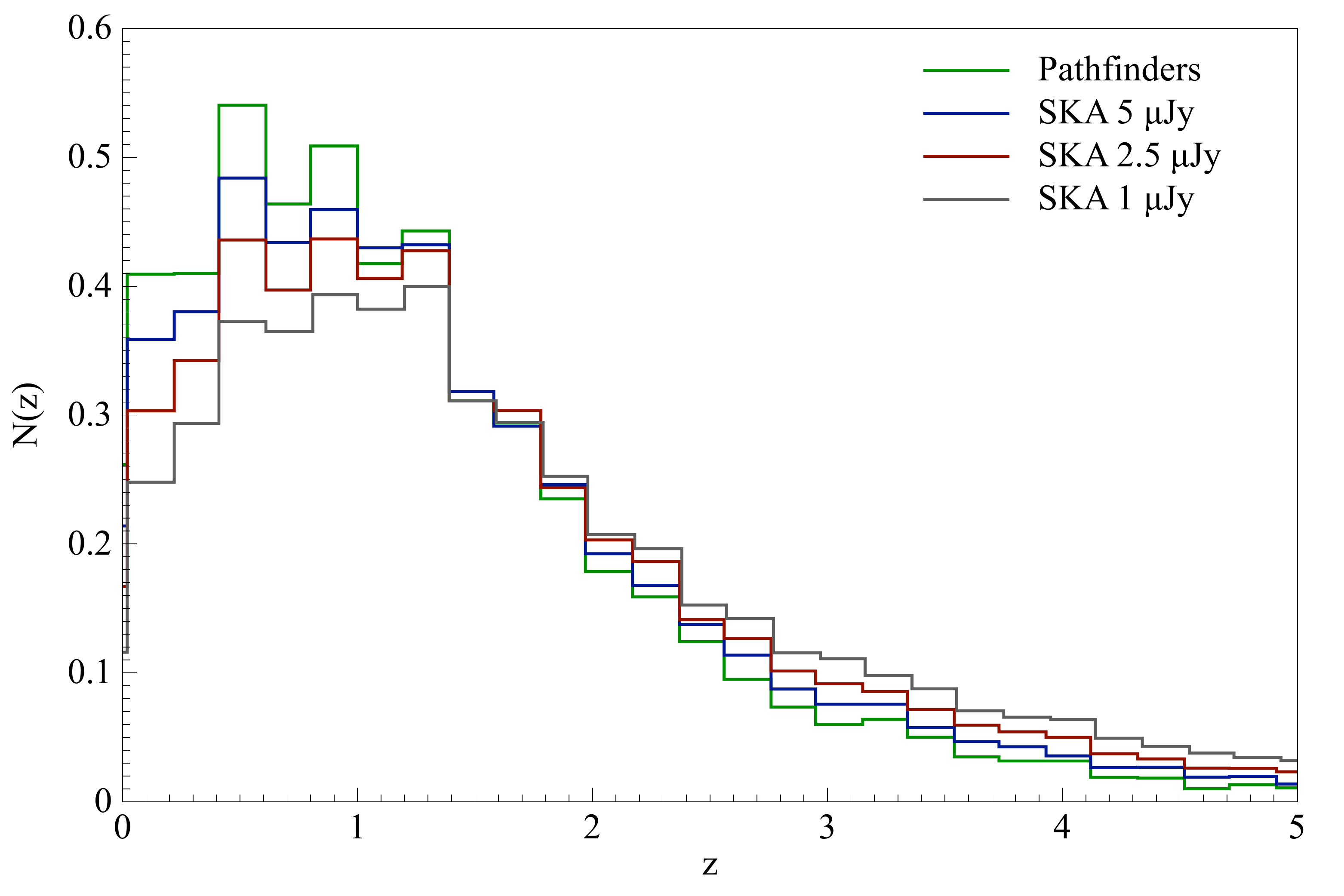}
\caption{Normalized redshift distributions for the SKA (in the three threshold limit considered), and the pathfinder surveys used in the paper (ASKAP and MeerKAT), assumed to have equal sensitivity.
}
\label{fig:Nz}
\end{figure*}

In order to make forecasts for cosmology using galaxy surveys, we need the number density of galaxies as well as the bias with which these galaxies trace the underlying dark matter. The number density of radio sources per redshift interval was extracted from the simulations in ~\cite{wilman08, wilman10} which are available through the $S^{3}$ database\footnote{http://s-cubed.physics.ox.ac.uk}.
These simulations provide prescriptions for the redshift evolution of the various populations which dominate the radio source counts.
We obtain distributions corresponding to the radio flux-density limits for the detection thresholds mentioned in Section~\ref{sec:surveys}; these are shown in Figure~\ref{fig:Nz}.

For the bias, we use the prescription described in ~\cite{wilman08, wilman10} for the $S^3$ simulation: each galaxy population is assigned a specific dark matter halo mass which is chosen to reflect the large-scale clustering found by observations; the bias is then simply the dark matter halo bias for that mass which is computed following the formalism in ~\cite{cole89, mo96}. For more details, see 
~\cite{raccanelliradio,ferramacho14}.
It is worth noting that large uncertainties in the modeling of the bias remain, although recent works~\cite{Lindsay14a, Lindsay14b} show that the low redshifts ($z<1$) the bias used for $S^3$ is in agreement with recent observation.
However, at high redshifts, the Fanaroff-Riley class I population, as defined in the SKADS simulation, may be more highly biased than previously assumed~\cite{Lindsay14a, Lindsay14b}. In our calculation we always marginalize over the amplitude of the bias (taken as a free parameter) for each redshift bin considered.


\subsection{Redshift information}
\label{sec:redshifts}
We explore effects of using redshift information of the radio continuum sources from cross-identification.
We should be in a position to identify the host galaxies of low-redshift radio sources using data from optical surveys; moreover, the iSW does not require high precision redshift information, so photometric redshifts are sufficient. Although current imaging surveys such as the Dark Energy Survey~\cite{bernstein12} and SkyMapper~\cite{skymapper} will provide some information, in the era of the SKA, LSST~\cite{lsst} will provide the key visible-wavelength data; more details about cross-identifications for SKA pathfinders can be found in~\cite{mcalpine12}. 
The increase in constraining power for the SKA-pathfinder generation given by including some degree of redshift information was investigated in~\cite{camera12}.

We make sharp cuts in the source redshift distribution, assuming that all the sources not identified by the optical surveys will be high-redshift radio sources. Although possibly not completely accurate, current data suggests that up to $\sim 90$~per cent of radio sources will be identified, with the vast majority of non-identifications occuring for radio sources at $z>2$~\cite{mcalpine12}.  In each configuration we therefore assume we have redshifts for all objects except those in the highest redshift bin.
Given that a number of surveys might be available in the next few years, here we explore different possibilities, where we divide the redshift distribution in different numbers of bins.
We study the following configurations:
\begin{enumerate}[i)]
\item 2 bins case: $0 < z < 1; \,\,\, 1 < z < 5$;
\item 3 bins low-z case: $0 < z < 0.5; \,\,\, 0.5 < z < 1; \,\,\, 1 < z < 5$;
\item 3 bins high-z case: $0 < z < 1; \,\,\, 1 < z < 2; \,\,\, 2 < z < 5$;
\item 5 bins case: $0 < z < 0.5; \,\,\, 0.5 < z < 1; \,\,\, 1 < z < 1.5; \,\,\, 1.5 < z < 2; \,\,\, 2 < z < 5$.
\end{enumerate}

In Section~\ref{sec:results} we present results obtained using these configurations. When assuming redshift information for our forecasted constraints, we then assume all the sources in the first $(n-1)$ bins are matched by optical counterparts, and so all the sources that are not cross-matched will be high-z radio sources. We will investigate the gain in constraining power given by the addition of a radio survey, i.e. the addition of the $n-th$ bin in each of the aforementioned cases.


\section{Fisher Analysis}
\label{sec:fisher}
In order to predict the precision in the measurement of \fnlt from the radio surveys we perform a Fisher matrix analysis~\cite{fisher35, tegmark97}; the Fisher matrix is given by:
\begin{equation}
F_{\alpha\beta}(z) =
\sum_{\ell=\ell_{\rm min}}^{\ell_{\rm max}} \frac{\partial {\bf C_\ell(z)}}{\partial \vartheta_\alpha}
{\sigma_{C_\ell(z)}^{-1}}\frac{\partial {\bf C_\ell(z)}}{\partial
\vartheta_\beta} {\sigma_{C_\ell(z)}^{-1}} \, , 
\label{eq:Fisher} 
\end{equation}
where $\vartheta_{\alpha(\beta)}$ is the $\alpha(\beta)$-th cosmological
parameter and $\sigma_{C_\ell}$ are errors in the correlation, defined as (see e.g.~\cite{cabre07}):
\begin{equation}
\label{eq:err-clgt}
\sigma_{C_{\ell}^{gT}} = \hksqrt{\frac{\left(C_{\ell}^{gT}\right)^2 + \left[ \left( C_{\ell}^{gg} + \frac{1}{\bar{n}_g} \right)C_{\ell}^{TT}\right]}{(2\ell+1)f_{\rm sky}}},
\end{equation}
where $f_{\rm sky}$ is the sky coverage of the survey and ${\bar{n}_g}$ is the average number density of sources.

To perform the Fisher analysis, we first parametrise our cosmology using:
\begin{equation}
\label{eq:paratriz} 
{\bf P} \equiv \{ w_{\rm 0}, w_{\rm a}, n_{\rm s}, \sigma_8, b_i, \Omega_{\rm m}, \Omega_{\rm \Lambda}, \Omega_{\rm K}, h, \gamma, f_{\rm NL} \} ,
\end{equation}
where $\{w_0,w_a\}$ are the parameters of the evolution of dynamical dark energy (see e.g.~\cite{linder03, linder05}), $n_{\rm s}$ is the primordial power spectrum index, $\{\Omega_{\rm m}, \Omega_{\rm \Lambda}, \Omega_{\rm K}\}$ are the energy density of matter, dark energy and curvature respectively, $\sigma_8$ is the overall normalisation (rms in spheres of radius $8 h^{-1} \rm Mpc$) of the matter density fluctuations, $b_i$ is the amplitude of the bias in each redshift bin, and $\gamma$ is the growth rate parameter describing the rate at which cosmological structures grow. We use the range $2 < \ell <200 $ and the Planck fiducial cosmology.

A complete analysis of the constraints on the cosmological parameters and the power in testing some specific models has been partially addressed in~\cite{raccanelliradio} and will be further investigated in a future work, which will also include other probes that will be available to radio surveys; here we are interested at the constraints on non-Gaussianity, so we marginalize over all the other parameters and show only constraints for $f_{\rm NL}$. This parameter is not very degenerate with the others, as it only affects the shape of the power spectrum at small wavenumber $k$.

We perform a full Fisher analysis and, to include the uncertainty on the product $b(z) \times N(z)$, we marginalize over the amplitude; when we consider the case with multiple redshift bins, we marginalize over the amplitude for every bin.


\section{Results}
\label{sec:results}
\subsection{No redshift information}
In this Section we compute the constraints on $f_{\rm NL}$ from iSW measurements assuming there is no redshift information for the radio sources.
To include the fact that we do not have a good knowledge of the bias and the amplitude of the radial number density, we assume a functional form for bias (the same used in~\cite{raccanelliradio}) and redshift distribution (from Figure~\ref{fig:Nz}), and then marginalize over an overall amplitude.

\begin{figure*}[htb!]
\includegraphics[width=0.7\linewidth]{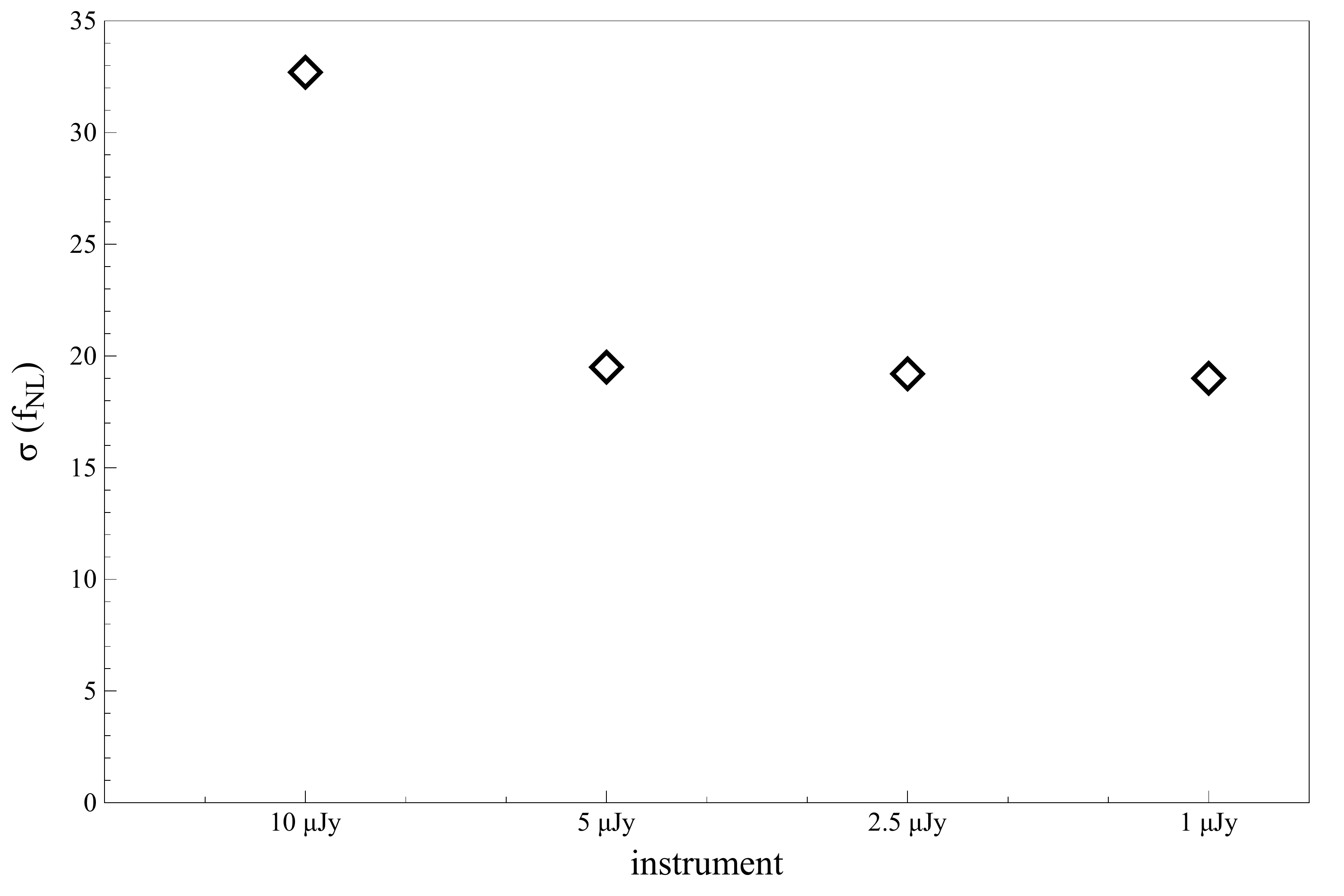}
\caption{Precision in the measurement of \fnlt for the radio experiments considered, for the case where no redshift information is used. The 10$\mu$Jy case is the expected sensitivity for the pathfinders ASKAP and MeerKAT, while the other three points are for different sensitivities of the SKA.}
\label{fig:emu1bin}
\end{figure*}

In Figure~\ref{fig:emu1bin} we show our results for the precision in the measurement of $f_{\rm NL}$ that it should be possible to obtain with the surveys considered; we show results for four different flux limits described in Section~\ref{sec:surveys}.
As one can see, there is a clear improvement in going from the flux-density limit possible with the precursor telescopes to the SKA (10 $\mu$Jy to 5 $\mu$Jy rms), but increasing the sensitivity further does not improve the measurement significantly.

\subsection{Including redshift Information}
In this Section we investigate the improvements that will be gained by combining the radio continuum surveys with optical and near-infrared data to obtain photometric redshifts, allowing us to divide our $N(z)$ into redshift bins.
We investigate the different redshift scenarios mentioned in Section~\ref{sec:redshifts}; in all cases, to account for the uncertainty in the knowledge of the redshift distribution and bias, we marginalize over an amplitude in each redshift bin.

\begin{figure*}[htb!]
\includegraphics[width=0.77\linewidth]{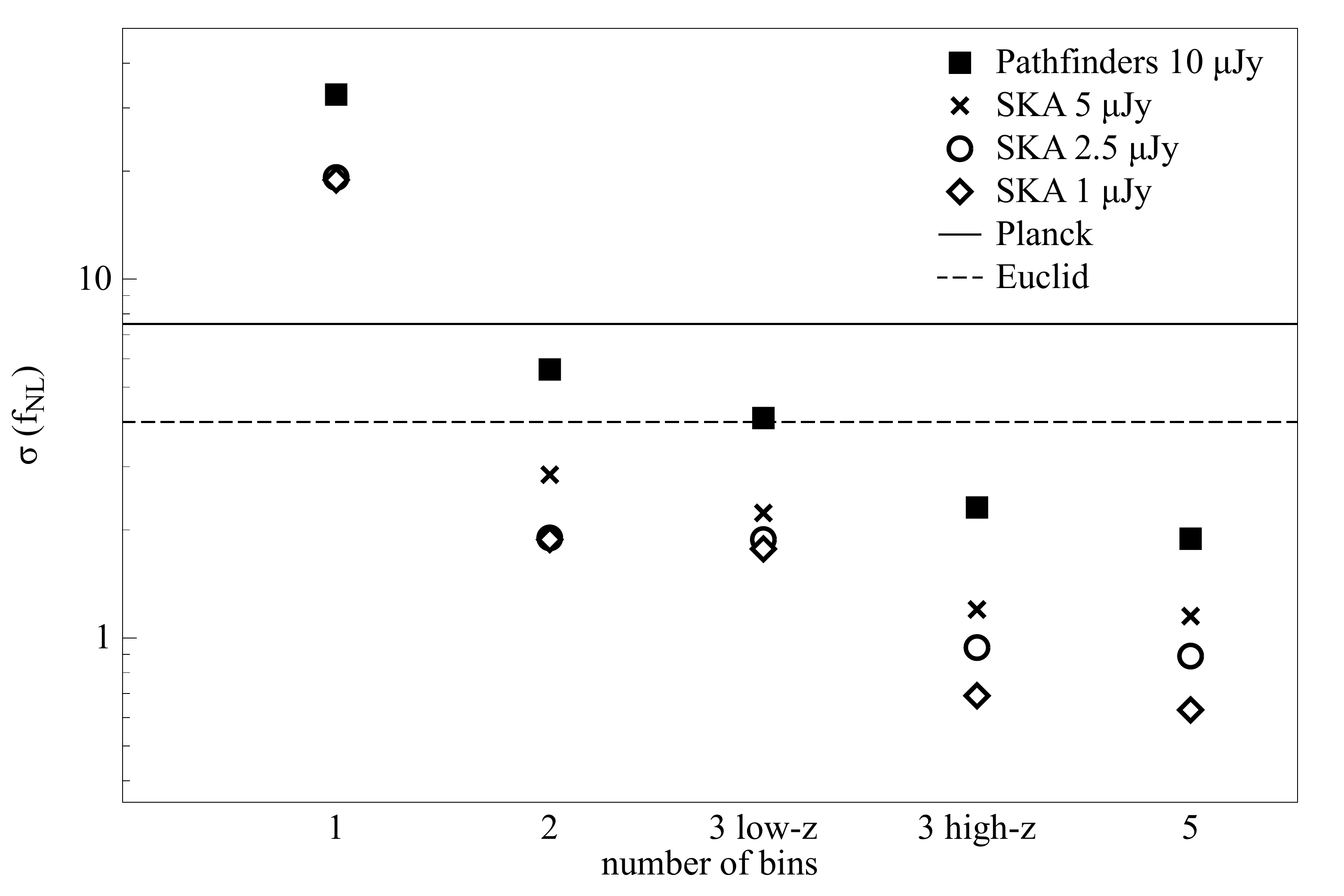}
\caption{
Effect of having different numbers of bins for the precision in measurements of the $f_{\rm NL}$ parameter, using the configurations of Section~\ref{sec:redshifts}, for the pathfinders instruments and the different flux limits of SKA.
}
\label{fig:zbins}
\end{figure*}

In Figure~\ref{fig:zbins} we show the constraints on $f_{\rm NL}$ for the different SKA surveys considered and for the different tomographic configurations, as a function of the number of bins, compared to current and predicted constraints coming from Planck and Euclid.
It can be seen that being able to divide the catalog into two redshift bins gives a huge improvement (of an order of magnitude), while dividing the sample into more bins gives only a moderate increase in constraining power (a factor of a few).
We consider two cases of 3 bins, as listed in Section~\ref{sec:redshifts}; one in the case where the maximum redshift for the cross-ID information is at $z=1$, and we call it ``low-z'', and one in which at $z=2$, and we call this ``high-z''.
The results show that a big improvement comes from having redshift information at high redshift.
The high-redshift information provides most of the constraining power, and once that is included, adding a larger number of bins does not bring a noticeable improvement in the constraining power.
This is useful for optimizing the science output of radio surveys: it is not really worth trying to increase the number of bins, but it would be very useful to have two bins and the ability to separate them.

\begin{figure*}[htb!]
\includegraphics[width=0.77\linewidth]{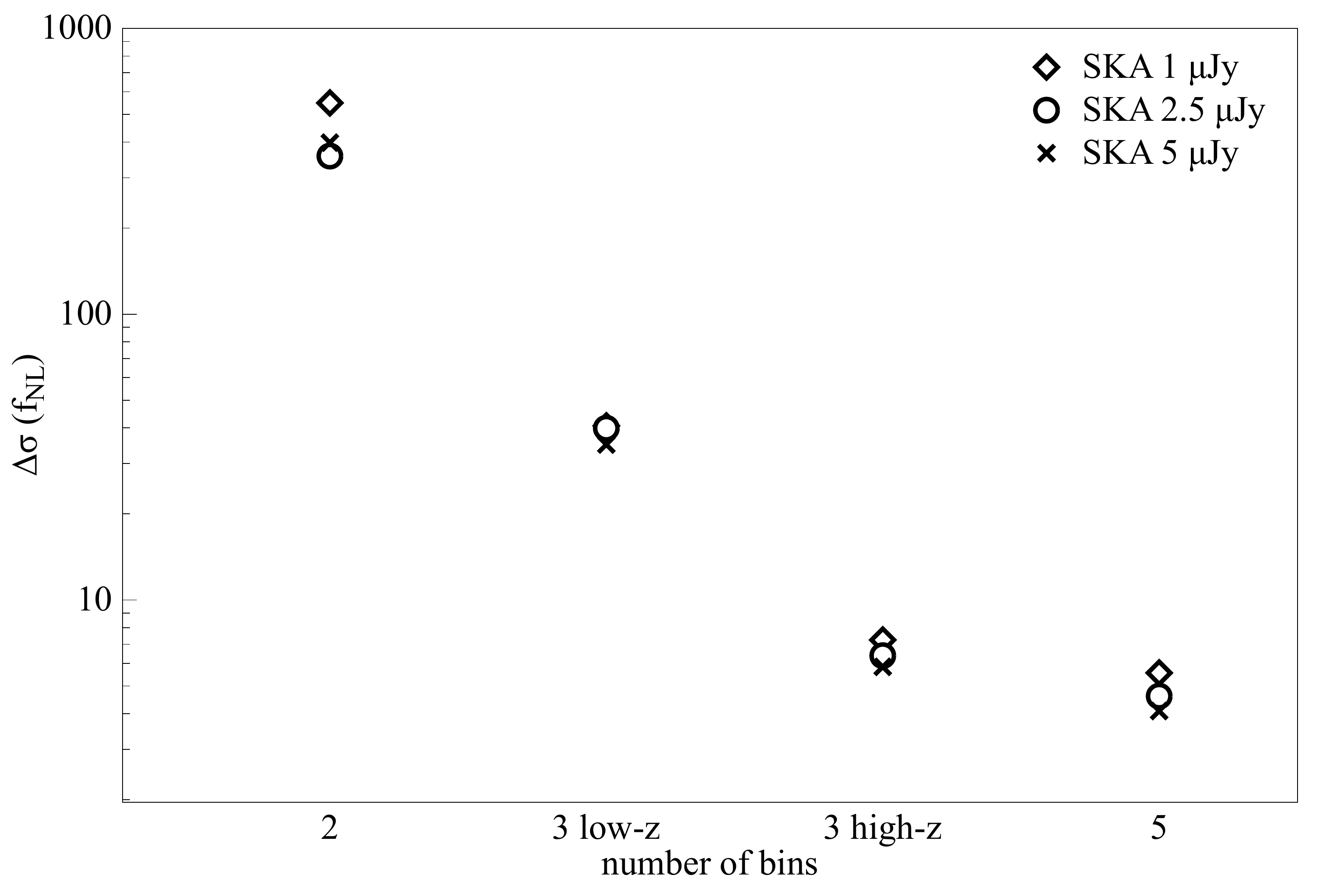}
\caption{
Effect of neglecting or including the radio high-z bin on the error on $f_{\rm NL}$ measurements.
}
\label{fig:noradio}
\end{figure*}

When we include the redshift information from optical imaging surveys, one could imagine that the analysis can be done using only an optical survey. We show that this is not the case, as the radio data provide an additional bin at high-z, reducing the error in the estimate of $f_{\rm NL}$. In Figure~\ref{fig:noradio} we show the improvements made by the addition of the radio high-z bin, plotting the ratio between the results using up to the $(n-1)th$ bin and the results using all the bins available (in practice adding radio). We define:
\begin{equation}
\Delta [\sigma(f_{\rm NL})] = \frac{\sigma(f_{\rm NL})_{\rm | \, (n-1) bins}}{\sigma(f_{\rm NL})_{\rm | \, n \, bins}} \, ,
\end{equation}
as the ratio of the error on the measurement of $f_{\rm NL}$ when considering only low-z sources that are matched with optical ones over the case including also the high-z bin from the radio survey.
This in practice indicates how much the constraining power is improved by adding the high-z bins coming from the radio survey.

We analyzed the case with redshift information for all three cases of different flux limits of SKA, and we divide the redshift distribution using the configurations of Section~\ref{sec:redshifts}, assuming we have the possibility to assign redshifts to radio sources via cross-ID.

In this section we predicted the constraints on $f_{\rm NL}$ for the different SKA and pathfinder surveys considered,  and for the different tomographic configurations. These radio surveys will provide superior constraints, compared to Planck or those predicted for Euclid, but only if some redshift information is available. Even just two redshift bins can reduce the error on $f_{\rm NL}$ by approximately an order of magnitude, and allow radio surveys to potentially provide some of the best measurements of $f_{\rm NL}$.


\section{Conclusions}
\label{sec:conclusions}
In this work we showed how we can use the integrated Sachs-Wolfe (iSW) effect from forthcoming radio surveys to probe the physics of the early universe, by searching for deviations from gaussianity in the initial conditions of the probability distribution function of cosmological perturbations.
We consider pathfinder-generation experiments (ASKAP and MeerKAT) and three possible configurations of the SKA, in the case with no redshift information, and when redshift information is available. 

Our results show that correlating the CMB with radio sources can give competitive constraints in the measurement of PNG parameters, in particular if we will be able to obtain information about the redshift of those sources. While, as expected, the single-bin case would provide constraints that are competitive with other LSS measurements, those are not competitive with CMB constraints.
In the case where redshift information is included, the large sky area surveyed by radio surveys coupled with the wide redshift-range will allow highly precise measurements of the non-Gaussianity parameter, with $\sigma (f_{\rm NL}) \lsim 1$ possible.
We also investigate the optimal configurations of the number density and redshift range of radio surveys in order to provide the best measurements and exploit to the maximum those experiments.

Therefore, future radio surveys, combined with future optical/near-infrared imaging surveys could provide very competitive constraints on \fnlt. For example, \cite{giannantonio12} predict that combining CMB data from Planck and galaxy data from a {\em Euclid}-like survey will reduce the 1$\sigma$ uncertainty on $f_{\rm NL}$ to $\approx 3$.
Considering long term opportunities, we found that measurements from the SKA and its combination with an optical survey that would allow a tomographic analysis should provide measurements close to the best that could be achieved (see also~\cite{RaccanellifNL}).

Finally, in this work we did not include a careful treatment of systematic errors; for some preliminary discussion on this, see~\cite{raccanelliradio}. More detailed analyses are currently being conducted.

\[\]{\bf Acknowledgments:}\\
AR would like to thank S. Stevanato for useful discussions.
Part of the research described in this paper was carried out at the  
Jet Propulsion Laboratory, California Institute of Technology, under a  
contract with the National Aeronautics and Space Administration.
DJB and RM were supported by the UK Science \& Technology Facilities Council (grant ST/K0090X/1).
RM, MGS and MJ were supported by the South Africa Square Kilometre Array Project and the South African National Research Foundation. 
MGS and SC acknowledges support from  FCT under grant PTDC/FIS-AST/2194/2012.
Parts of this research were supported by the Australian Research Council Centre of Excellence in All-sky Astrophysics (CAASTRO), through project number CE110001020.


%
%
\end{document}